\documentclass[aps,prl,twocolumn,groupedaddress,superscriptaddress]{revtex4}

\usepackage{graphicx}
\newcommand{\vect}[1]{{\mbox{\boldmath $#1$}}}
\newcommand{\ket}[1]{\vert #1 \rangle}
\newcommand{\bra}[1]{\langle #1 \vert}
\newcommand{\eqref}[1]{Eq.~(\ref{#1})}
\newcommand{\figref}[1]{Fig.~\ref{fig:#1}}

\newcommand{\deffig}[4]{
\begin{figure}[tb]
  \begin{center}
  \includegraphics[width=#3 \textwidth]{#2}
  \end{center}
  \caption{ \label{fig:#1} #4}
\end{figure}
}
\begin{document}
\title{N\'eel and Spin-Peierls ground states of two-dimensional SU($N$)
  quantum antiferromagnets
}
\author{Kenji Harada}
\affiliation{
  Department of Applied Analysis and Complex Dynamical Systems,
  Kyoto University, Kyoto 606-8501, Japan
}
\author{Naoki Kawashima}
\affiliation{
  Department of Physics, Tokyo Metropolitan University,
  Tokyo 192-0397, Japan
}
\author{Matthias Troyer}
\affiliation{
  Theoretische Physik, ETH Z\"urich, CH-8093 Z\"urich, Switzerland
}
\date{\today}
\begin{abstract}
The two-dimensional SU($N$) quantum antiferromagnet,
a generalization of the quantum Heisenberg model, is
investigated by quantum Monte Carlo simulations.
The ground state for $N\le 4$ is found to be of the N\'eel type
with broken SU($N$) symmetry,
whereas it is of the Spin-Peierls type for $N\ge 5$
with broken lattice translational invariance.
No intermediate spin-liquid phase was observed
in contrast to previous numerical simulations on smaller lattices
[Santoro et al., Phys.\ Rev.\ Lett.\ {\bf 83} 3065 (1999)].

\end{abstract}
\maketitle
%

%
%
The existence of a short range resonating valence bond (RVB)
spin liquid \cite{Anderson1973} is one of the central problems
for low-dimensional quantum spin systems. An RVB spin liquid
exhibits a finite gap for spin excitations,
has only short range order, and does not break any lattice symmetry. The
search for RVB spin liquid states was motivated by the
suggestion that
such strongly-correlated but quantum disordered states can be
turned into a super conducting state upon doping,
which may explain the mechanism of the copper-oxide super conductors
\cite{Anderson1987}.

An RVB spin liquid is presumably created by strong quantum fluctuations
which destroy magnetic ordering. The simplest construction of spin
liquid states is on lattices with an even number of spins per unit cell,
such as two spin ladders \cite{ladders}, bilayer\cite{bilayer} or coupled
plaquettes\cite{plaquette}. There a strong coupling within the unit cells
leads to the formation of weakly coupled spin singlets.

This mechanism does not apply to lattices with an odd number of spins per
unit cell, such as the square lattice relevant for the high-$T_c$ cuprates.
In the square lattice Heisenberg antiferromagnet quantum fluctuations
decrease the N\'eel order as the spin $S$ is decreased whereas
the ground state
remains ordered even for $S=1/2$ \cite{Manousakis}.
Stronger quantum fluctuations are thus needed and frustrating interactions
have been proposed as one route \cite{FazekasA1974}.
Since frustrating interactions generally cause a sign problem for quantum
Monte Carlo simulations, numerical calculations are usually restricted to
small lattices and it is hard to draw definitive conclusions. Still, it
could be established that the ground state of the antiferromagnetic
Heisenberg model on a triangular lattice is magnetically ordered
\cite{BernuLP1992,CapriottiTS1999}.
A disordered ground state was suggested
\cite{WaldtmannEtal1998}
for the fully frustrated model on a kagom\'e lattice, but the nature of this
state is not clear. The only clear evidence for an RVB spin liquid in a
frustrated system so far is for the hardcore dimer model on a triangular
lattice \cite{dimer1,dimer2}.

The route we follow in this Letter is to increase quantum fluctuations by
considering models with higher symmetry than SU(2) and determine the nature
of ground state once quantum fluctuations are strong enough to destroy
N\'eel order.  While SU(3) symmetric points occur in spin $S=1$ Heisenberg
antiferromagnets with additional biquadratic interactions
\cite{HaradaK2001,HaradaK2002,BatistaOG2002}, SU(4) symmetric models have
had a long history as special points in coupled spin-orbital models
\cite{KugelK1973}. Quantum Monte Carlo simulations for the SU(4)-symmetric
Kugel-Khomskii model \cite{KugelK1973} suffer from a negative sign problem
in more than one dimension \cite{Frischmuth} and numerical simulations are
thus restricted to small lattices \cite{Bossche}. For a related
antiferromagnetic SU(4) model  in which there is no negative sign problem,
an RVB spin liquid ground state was suggested  by previous quantum Monte
Carlo simulations \cite{SantoroSGPT1999}, as the first example of such a
state on a non-frustrated lattice.
These models can be regarded as a generalized Heisenberg
antiferromagnet, and belong to a family of SU($N$) models employing one of
the simplest representations for the symmetry group of the model. In this
sense they are ``minimal'' models for
physical systems with higher symmetries than SU(2),
and the results are expected to be of relevance as reference models in a
wide array of applications.

An SU($N$) model is specified by an irreducible representation of the
operators,
i.e., the Young tableau for the SU($N$) algebra.
Here we consider the series of models with a single-row Young tableau.
In one dimension the ground state of these models is
dimerized for any fixed number of columns $n_{\rm c}$
of the Young tableau as $N\rightarrow\infty$ \cite{Affleck1985}.
In two dimensions, in a bosonic representation equivalent to a single-row
Young tableau,
it was found that for large $n_{\rm c}$ and $N$,
the ground state is N\'eel state if $N < \kappa n_{\rm c}$,
where $\kappa$ is a numerical constant \cite{ArovasA1988}.
Read and Sachdev \cite{ReadS1990} confirmed this result and,
by examining the continuous version of the model in
the large-$N$ limit,
argued that the ground state becomes a spin-Peierls state
as soon as $N$ exceeds the boundary $N^{\ast}\equiv \kappa n_{\rm c}$.
Although their approach is less reliable for small $N$,
they conjectured a direct transition with no spin liquid intermediate phase
even for small values of $N$ (i.e., for small $n_{\rm c}$) -- inconsistent
with the numerical observation of a spin liquid phase in the SU(4) model
\cite{SantoroSGPT1999}.
In this Letter, using a new cluster quantum Monte Carlo algorithm,
we present conclusive numerical evidence for a direct transition
from the N\'eel state to the spin-Peierls state.
No RVB spin liquid ground state is realized for any $N$, in contrast to
the previous simulations on small lattices \cite{SantoroSGPT1999}.

Our SU($N$)-invariant generalization of
the Heisenberg model\cite{Affleck1985,ReadS1990} can be formally written as
\begin{equation}
  \label{eq:hamiltonian}
  H = \sum_{\langle \vect{r}\vect{r}' \rangle} H_{\vect{r}\vect{r}'}
    = \frac{J}{N} \sum_{\langle \vect{r}\vect{r}' \rangle}
  J^{\alpha}_{\beta}(\vect{r}) J^{\beta}_{\alpha}(\vect{r}'),
\end{equation}
where
$\langle \vect{r}\vect{r}' \rangle$ runs over all nearest-neighbor pairs
on a square lattice,
and repeated indices $\alpha, \beta = 1,\dots,N$ are
to be summed over.
Throughout this letter, we set $J=1$ as the unit of the energy.
The symbols $J^{\alpha}_{\beta}(\vect{r})$
denote the generators of SU($N$) algebra, that satisfy
$
  [J^{\alpha}_{\beta}(\vect{r}),J^{\mu}_{\nu}(\vect{r}')]
  =
  \delta_{\vect{r},\vect{r}'}
  \left(
  \delta^{\alpha}_{\nu} J^{\mu}_{\beta}(\vect{r})
  -
  \delta^{\mu}_{\beta}  J^{\alpha}_{\nu}(\vect{r})
  \right).
$
To uniquely  specify the model, we have to choose the
representation of the algebra.
The model examined in this letter is the ``antiferromagnetic''
SU($N$) model, in which
an operator $J^{\alpha}_{\beta}$ is represented by a $N$
dimensional matrix with the fundamental representation
(i.e., a single-box Young tableau) on one sublattice and with the
conjugate representation (i.e., the single-column Young tableau
with $N-1$ boxes) on the other sublattice. We note that the Kugel-Khomskii
model \cite{KugelK1973} uses the same representation on both sublattices and
that
for $N=2$ both models reduce to the ordinary SU(2) antiferromagnetic
Heisenberg model.

In this representation, the model can be conveniently
expressed in terms of SU(2) spins with $S=(N-1)/2$.
The matrix elements of the pair Hamiltonian $H_{\vect{r}\vect{r}'}$
are explicitly given by
\begin{equation}
  \label{eq:spin-hamiltonian}
  \bra{\alpha^\prime, \beta^\prime} H_{\vect{r}\vect{r}'}
  \ket{\alpha,\beta} = -
  \frac{J}{N}  \   \delta_{\alpha, -\beta}
  \  \delta_{\alpha^\prime,-\beta^\prime},
\end{equation}
where $\ket{\alpha, \beta}$ ($\alpha, \beta= -S, -S+1, \cdots, S$)
is the simultaneous eigenstates of the $z$ components of
SU(2) spin operators, $S^z(\vect{r})$ and $S^z(\vect{r}')$.
We probe for two types of long-range order. The first is a generalized
N\'eel state with broken SU($N$) symmetry characterized by
a finite staggered magnetization
\begin{equation}
  M_{\rm s} \equiv \sum_{\vect{r}} (-1)^{\vect{r}} S^z(\vect{r}).
\end{equation}
The second is a dimerized state with broken translational invariance but no
broken spin rotation symmetry, characterized by an order parameter such as
\begin{equation}
  D_{\vect{k}} \equiv \sum_{\vect{r}} e^{-i\vect{k}\vect{r}}
  S^z(\vect{r}) S^z(\vect{r}+\vect{e}_x)
\end{equation}
where $\vect{e}_x$ is the lattice unit vector in the $x$ direction.
It was argued \cite{ReadS1990} that
$\vect{k} = (\pi,0)$ or $(0,\pi)$ is preferred
when the lattice translational symmetry is broken.

Recent developments in quantum Monte Carlo algorithms \cite{LoopAlgorithm}
allow us to perform quantum
Monte Carlo (QMC) simulations on larger systems and for a wider range of
$N$
than was possible in previous calculations \cite{SantoroSGPT1999}.
By splitting each original spin operator into $2S$ Pauli spins with $S=1/2$
\cite{KawashimaG1994},
the Hamiltonian of the SU($N$) model
in the new extended Hilbert space is expressed
\cite{HaradaK2001} as
\begin{equation}
  H_{\vect{r}\vect{r}'} = -\frac{J}{N}
  \left[ \Delta_{\rm Horizontal}^{(N-1)}(\vect{r},\vect{r}') \right]_s,
  \label{eq:symmetrized}
\end{equation}
where the symbol $[\cdots]_s$ denotes the symmetrization
with respect to the $2S$ Pauli spins.
The symbol $\Delta_{\rm Horizontal}^{(N-1)}$ denotes the operator
whose matrix element is one if the $(N-1)$-fold horizontal
graph matches the initial and the final spin states,
while it is 0 otherwise (See \figref{horizontal}).
\deffig{horizontal}{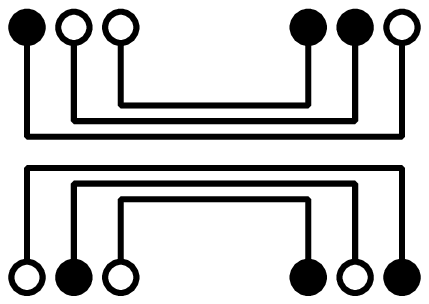}{0.16}
{One of the three-fold horizontal graphs for $N=4$ corresponding to
$\Delta^{(3)}_{\rm horizontal}(\vect{r},\vect{r}')$,
with one of its matching spin-configurations.
(A different choice of the graph does not make any difference
because of the symmetrization in \eqref{eq:symmetrized}.)
Open circles denote up-spins and filled circles down-spins.
A spin-configuration matches a graph if and only if
any two connected spins are antiparallel.
}
This formulation yields the following quantum Monte Carlo algorithm:
(i) for a given world-line configuration,
distribute the $(N-1)$-fold horizontal graphs
with the density $J/N$,
(ii) construct loops by following world lines and horizontal graphs, and
(iii) flip each loop independently with probability 1/2.
The majority of results reported in this Letter were obtained
with this algorithm.
Better algorithms could be constructed recently
using the framework of the stochastic series expansion \cite{SSE}
and the idea of coarse-graining \cite{HaradaK2002b},
or directly working with the weight equation \cite{KawashimaG1995}
in the conventional framework of loop algorithms.
This latter algorithm was used in a part of the calculations.
The details of these algorithms will be reported
elsewhere \cite{HaradaKfuture}.

%
%
Simulations have been performed at low enough temperatures to be effectively
in the ground state; careful checks have been performed for each system size
and value of $N$ by varying the temperature. We explored system sizes $L$ up
to $L=128$ for $N=3$ and $4$,
and $L=64$ for $N=5,6,7$ and $8$.
The number of Monte Carlo sweeps in a typical set of simulation is
$3.4 \times 10^5$ for the most time consuming case ($L=128$ and $N=4$).

In Fig. \ref{fig:ss-all} we show the spin structure factor,
$ S_{(\pi,\pi)} = L^{-d}\langle (M_{\rm s})^2 \rangle $,
divided by $L^2$, which in the SU($N$) language reads
$
  S_{(\pi,\pi)} =
  \frac{N^2(N+1)}{12}\left\langle
  \left( \sum_{\vect{r}} J^1_1(\vect{r}) \right)^2
  \right\rangle.
$
\deffig{ss-all}{fig2.eps}{0.45}
{Static structure factors $S_{(\pi,\pi)}$ for $2 \le N \le 8$.
The straight line representing the power law,
$S_{(\pi,\pi)}L^{-2} \propto L^{-2}$, is drawn for comparison.
Estimated statistical errors are not shown because
they are equal to or smaller than the symbol size.
The inset presents the data for $N=3,4,5,6$ in the linear scale,
together with the best fitting curves obtained
by the method of least-squares.
}
In the N\'eel phase,
$S_{(\pi,\pi)}/L^2$ converges
to the square of the staggered
magnetization per spin, $m_s$, as $L\to\infty$, while it decreases
asymptotically to zero, being proportional to $L^{-2}$,
in the absence of N\'eel order. Our results
in \figref{ss-all} show clear evidence for N\'eel order for $N\le4$ and
absence of N\'eel order for $N\ge5$,
indicating that
the phase boundary of the N\'eel ground state
lies between $N=4$ and $5$.

For $N\le4$ the staggered magnetization
in the thermodynamic limit is calculated from a finite size extrapolation
using a second order polynomial in $1/L$.
A least-squares fit based on the data for $L\ge 8$,
gives
\begin{eqnarray}
  m_s & = & 0.1682(6) \qquad \mbox{for $N=3$}, \\
  m_s & = & 0.091(3) \qquad \mbox{for $N=4$}
\end{eqnarray}
where the numbers in the parentheses indicate the estimated statistical
error
(of one standard deviation) in the last digit.
For $N=5$ and $N=6$, the estimated $m_s^2$ agrees with
zero within the statistical error ($m_s^2 = -0.0003(6)$ for $N=5$
and $m_s^2 = 0.0004(5)$ for $N=6$).

The existence of N\'eel order at $N=4$ is further confirmed by
the correlation ratio, which eliminates contributions from short-range
correlations \cite{TomitaO2002}.
While the structure factor $S_{(\pi,\pi)}$ is a sum of two-point
correlation functions with all distances and therefore contains
short-range correlations, the quantity
$$
  \Gamma_{M_{\rm s}}(\tau)
  \equiv
  \langle M_s(\tau)M_s(0) \rangle
  \equiv
  \langle e^{\tau H} M_s e^{-\tau H} M_s \rangle
$$
does not if $\tau$ is sufficiently large.
We measure this quantity at a fixed aspect ratio,
$\beta/L$, and compute the ratio
$\Gamma_{M_{\rm s}}(a'\beta\,|\,L,\beta)
/\Gamma_{M_{\rm s}}(a\beta\,|\,L,\beta)$
with $a<a'$ for various system sizes.
As $L$ is increased, this ratio converges to 1
if the system has long range order.
If, on the other hand, the system is disordered
with a finite gap and therefore has a finite correlation
length in the imaginary-time direction,
it will converge to 0.
Hence this ratio serves as a good indicator of the existence of
a long range order, similar to the well-known Binder parameter.
In \figref{ss-c}, we plot the correlation ratio
for various aspect ratios $\beta/L=1.0, 1.5$ and $2.0$ for $N=4$.
We can see that the correlation ratio approaches 1 and thus establish
long range N\'eel order.

\deffig{ss-c}{fig3.eps}{0.35}{The ratio of dynamic correlations
of staggered magnetization at two imaginary-time intervals,
$\beta/2$ and $\beta/4$, for the SU(4) model.
As $L$ increases, the value approaches 1,
regardless of the aspect ratio $\beta/L$.
}

The disagreement of our conclusion (N\'eel order for $N=4$) with that of
Ref.  \cite{SantoroSGPT1999} is not in the raw numerical data. Their
estimates of $S_{(\pi,\pi)}$ agree with ours
shown in \figref{ss-all} within the statistical errors.
The disagreement is solely due to the small system sizes studied in Ref.
\cite{SantoroSGPT1999}.
The convergence of the magnetization to a finite value in \figref{ss-all}
can be seen only for system sizes $L\ge 32$,
whereas the previous simulations were limited to $L = 12$

In order to fully answer the question whether an intermediate
spin-liquid phase exists, we next determine at which value of $N$
the ground state starts being dimerized.
Figure \ref{fig:sd2} shows the
dimer structure factor,
$ S^{D}_{\vect{k}} \equiv L^{-d} \langle D_{-\vect{k}}D_{\vect{k}} \rangle$,
for $\vect{k} = (\pi,0)$
divided by $L^2$ at $N=4,5$ and $6$.
In the thermodynamic limit,
$S^D_{\vect{k}}/L^d$ should converge to the squared dimerization per spin.
In \figref{sd2} we see a clear power law decay of $S^D_{\vect{k}}/L^2$
following $L^{-2}$ and thus absence of dimerization for $N=4$, but a slower
decay and up-ward bending for $N=5$ and $6$.
This suggests convergence to finite values,
although the examined systems are not large enough
to cover the region where $S^D_{(\pi,0)}$ shows no size dependence.
\deffig{sd2}{fig4.eps}{0.45}{
The ${\vect{k}}=(\pi,0)$ dimer structure factors $S^D_{(\pi,0)}$
for $N=4, 5, 6$ in logarithmic scale.
The inset is the linear-scale plot.
The solid lines in the inset are the best fitting curves of
least-squares based on the $L \ge 8$ data.
}
For a more systematic analysis,
we once more perform a  least-squares fit
of the data of $S^D_{(\pi,0)}$ for $L \ge 8$, using
a second order polynomial in $1/L$.
Our results for the spontaneous dimerization per site
are 0.103(3) for $N=5$
and 0.18(5) for $N=6$ (See the inset of \figref{sd2}).
For $N=4$, the same analysis yields
$ (D_{(\pi,0)}/L^d)^2 = -0.00002(2) $,
consistent with absence of dimerization.

We also compute $S^D_{\vect{k}}$ at $\vect{k} = (\pi,\pi)$
and find $S^D_{(\pi,\pi)}/L^2 \propto L^{-2}$,
for $N = 4,5$ and $6$, indicating absence of dimerization
at this wave vector, consistent with the
previous suggestion\cite{ReadS1990}.


%
%

In conclusion, our high-accuracy QMC simulations using new loop-type QMC
algorithms
have shown that the ground state of the SU($N$) square lattice
antiferromagnet
 is the N\'eel state  for $N\le 4$,
whereas it is the dimerized or Spin-Peierls state for $N \ge 5$.
No intermediate spin-liquid has been observed,
consistent with analytical arguments \cite{ReadS1990} but inconsistent with
previous numerical simulations on smaller lattices \cite{SantoroSGPT1999}.

It is interesting to compare the present result with the
analytical estimate of the phase boundary \cite{ArovasA1988}:
$n_{\rm c} \sim 0.19 N^{\ast}$
where $n_{\rm c}$ is the number of the columns of the Young tableau.
This estimate is supposed to be accurate for large $N$,
and is not necessarily justified for the present case where
$n_{\rm c} = 1$ but is still surprisingly accurate. For our model it would
indicate N\'eel order up to $N=5$, while we find N\'eel order only up to
$N=4$.

Concerning the existence of true RVB spin liquid states without any broken
symmetry, our results are essentially negative. By showing that the proposed
spin liquid state in an SU(4) model exhibits N\'eel order,
we are left with
only the hardcore dimer model on a triangular lattice \cite{dimer1} as a
model with clearly established spin liquid ground state
\cite{dimer1,dimer2}. Since the hardcore dimer model on the square lattice
does not show any gapped spin liquid phase \cite{RK}, and in numerical
simulations only symmetry broken phases were found so far for frustrated
lattices,  we are led to conjecture that a spin liquid state without any
broken symmetry seems to be impossible to obtain on a bipartite lattice with
an odd number of spins in the unit cell.

%
%
\begin{acknowledgments}
We thank C. Batista, S. Sorella and F.C. Zhang for useful comments and
suggestions.
Most of the numerical calculations for the present work have
been performed on the Hitachi SR-2201 on the Institute for
Solid State Physics, University of Tokyo.
The present work is supported by Grants-in-Aid for Scientific
Research Program (\# 12740232 and \# 14540361) from Monkasho, Japan
and by the Swiss National Science Foundation.
\end{acknowledgments}
\end{document}